\documentclass{article}
\usepackage[table]{xcolor}
\usepackage[utf8]{inputenc}

\usepackage{tabularx}
\usepackage{longtable}

\usepackage[T1]{fontenc}
\usepackage{amsmath,amssymb,amsthm,mathtools}
\usepackage{paralist}

\usepackage{xr-hyper}
\usepackage{listings}
\usepackage{listings}
\usepackage{xcolor}

\definecolor{dkgreen}{rgb}{0,0.6,0}
\definecolor{gray}{rgb}{0.5,0.5,0.5}
\definecolor{mauve}{rgb}{0.58,0,0.82}

\usepackage{graphicx, caption}
\usepackage{amsfonts}
\usepackage{mathtools}
\usepackage{dsfont}
\usepackage[a4paper,top=2cm,bottom=2cm,left=3.2cm,right=3.2cm]{geometry}
\usepackage{array}
\usepackage[colorinlistoftodos]{todonotes}
\usepackage[colorlinks=true, allcolors=blue]{hyperref}
\usepackage{placeins}
\usepackage{lineno}
\usepackage[parfill]{parskip} 

\newcolumntype{L}[1]{>{\raggedright\let\newline\\\arraybackslash\hspace{0pt}}p{#1}}
\newcolumntype{C}[1]{>{\centering\let\newline\\\arraybackslash\hspace{0pt}}p{#1}}
\newcolumntype{R}[1]{>{\raggedleft\let\newline\\\arraybackslash\hspace{0pt}}p{#1}}

\definecolor{growth}{HTML}{228b22}

\definecolor{replication}{HTML}{6699CC}
\definecolor{death}{HTML}{888888}
\definecolor{displacement}{HTML}{DDCC77}
\definecolor{conjugation}{HTML}{117733}
\definecolor{co-conjugation}{HTML}{999933}
\definecolor{loss}{HTML}{882255}
\definecolor{partial loss}{HTML}{CC6677}

\graphicspath{{Figures/}}
\usepackage[style=numeric, citestyle=numeric-comp, sorting=none]{biblatex}
\title{A guide to modeling plasmid co-infection dynamics}
\date{}

\author{Berit Siedentop$^1$$^*$$^\dagger$, Jana S. Huisman$^2$$^*$, Claudia Igler$^3$\\
\footnotesize $^1$ Department of Infectious Diseases and Hospital Epidemiology, University Hospital Zurich and University of Zurich, Zurich, Switzerland \\
\footnotesize $^2$ Physics of Living Systems, Massachusetts Institute of Technology, Cambridge, MA, United States of America \\
\footnotesize $^3$ Division of Evolution, Infection and Genomics, School of Biological Sciences, University of Manchester, Manchester, UK \\
\footnotesize $^*$ Authors contributed equally \\
\footnotesize $^\dagger$ Correspondence: berit.siedentop@uzh.ch}

\begin{document}
\maketitle 

\begin{abstract}
    Mobile genetic elements like conjugative plasmids play a crucial role in shaping the genetic content and population dynamics of bacterial species. Bacterial populations often contain not one, but multiple co-circulating MGEs, which modify each other's population dynamics in a myriad of ways. Mathematical modeling is a powerful tool to gain intuition about the expected eco-evolutionary dynamics in a biological system with many interacting players. Here, we detail how to develop a mathematical model of plasmid co-infection, how to implement this computationally, and we give examples of what can be learned from such models.
\end{abstract}

\section{Introduction}

Mobile genetic elements, such as plasmids, play a crucial role in shaping the genetic content and population dynamics of bacterial species. Bacterial populations often contain not one, but multiple co-circulating plasmids, which modify each other's population dynamics in complex ways~\cite{igler2022}. Co-infecting MGEs can prevent the entry, replication, or transmission of other MGEs. There can be epistatic effects on fitness costs and benefits for the host, as well increasing rates of transmission due to co-transfer. All these different processes come together in highly complex and nonlinear ways, making it hard to predict what the effect of a small change in one of these processes may do to the population dynamics of the MGEs as a whole. 

Mathematical modeling is a powerful tool to understand the population dynamics of biological systems with many interacting players. Modeling can be used to connect biological mechanisms, such as toxin-antitoxin systems or superinfection exclusion, to population-level outcomes like the prevalence of a plasmid or the invasibility of a population. Modeling can also be used to determine which biological trait is most important to determine population level outcomes in a particular environment, or to explore the interactions between different traits. Contrary to experiments, modeling allows for cheap and fast comparison of a range of different biological traits and systems, which can generate hypotheses and identify interesting parameter sets for experimental manipulation. These predictions can then be tested with experiments or bioinformatics. 

Some examples of questions about plasmid co-infection dynamics that have been studied with the help of mathematical models are: how non-conjugative plasmids can be maintained~\cite{Werisch2017,Levin1980}, when conjugative plasmids can coexist~\cite{Hoeven1985, Lopatkin2017,Gregory2008,Gama2020}, which factors influence gene mobility between plasmids~\cite{Condit1990} or plasmids and the chromosome~\cite{Lehtinen2021}; and how traits like toxin-antitoxin systems~\cite{Mongold1992} and surface exclusion~\cite{Hoeven1985} evolve. 

Models fall into many different classes: deterministic or stochastic, agent- or population-based, simple or complex. The choice of model depends on the specific question and system being studied, as each type offers distinct advantages and disadvantages in terms of complexity, realism, computational efficiency, and interpretability (see section \ref{sec:model_type_considerations}). 
Here, we detail how to build a general deterministic model of plasmid co-infection, starting from a single plasmid infection model. We explain how to implement these models computationally, and how they can be used to qualitatively understand the relationship between different biological parameters and eco-evolutionary outcomes.


\section{Materials} \label{sec:materials}

To run simulations of mathematical models computationally, they have to be implemented in a programming language. This requires installation of that language (e.g.  R \cite{R_language}) and, if not included, an environment to edit code (e.g. RStudio \cite{Rstudio}).
For implementation in R we use the deSolve package \cite{deSolve}, which provides numerical solutions of the developed ordinary differential equation model, and for analysis and visualization we use base R (packages like tidyverse \cite{R_tidyverse} and ggplot2 \cite{R_ggplot2} can be used for more advanced analysis and visualization). 

A free guide on how to install R, Rstudio and specific packages in R, as well as how to take the first steps in R, is provided, for example, by \textit{Wickham and Grolemund 2017} \cite{wickham2017}.

We provide R code for numerical simulations of the plasmid models in section \ref{full_R_code}; however, numerical model output can also be generated with other common coding languages such as Python \cite{python}, Julia \cite{julia}, or Matlab \cite{MATLAB}. 

\section{Method}

\subsection{General considerations about model type} 
\label{sec:model_type_considerations}

\begin{figure}[ht!]
\centering
\includegraphics[width=0.7\linewidth]{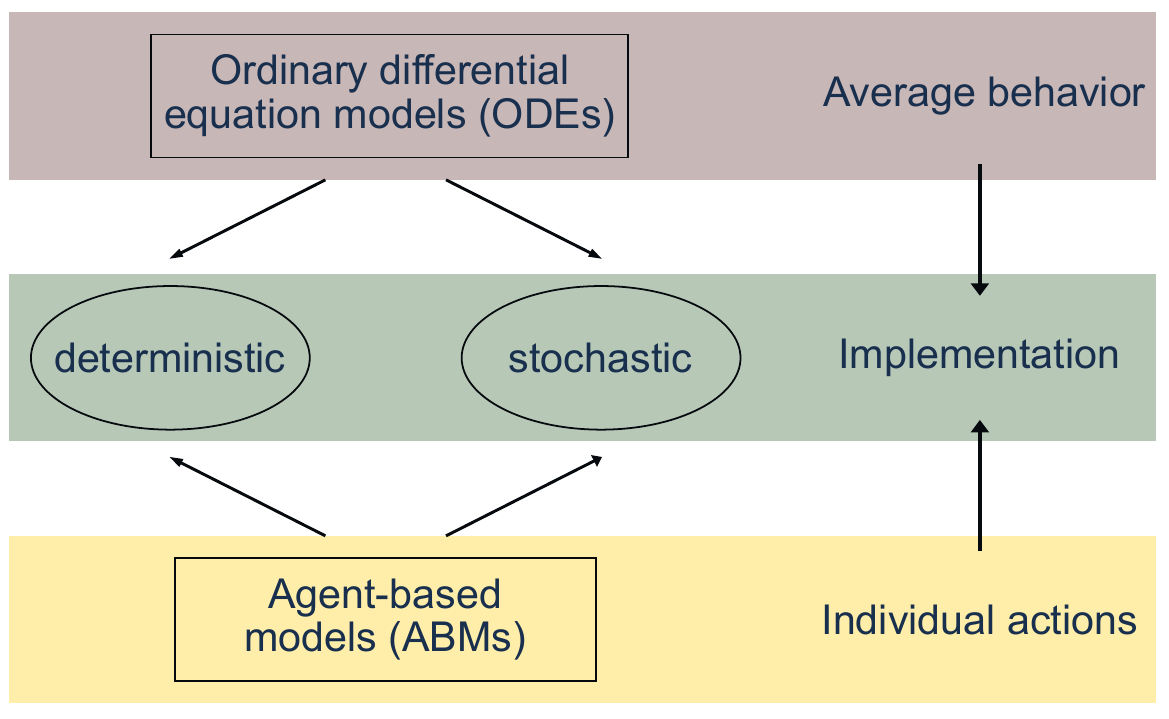}
\caption{\textbf{Schematic illustrating different model and implementation types.} } \label{fig:schematic_model_type}
\end{figure}

Different types and implementations of mathematical models can be used to describe biological processes that determine plasmid dynamics such as plasmid conjugation (Fig.~\ref{fig:schematic_model_type}). These models can be highly complex, describing every mechanistic step involved in plasmid conjugation (e.g. gene expression, pilus formation and attachment, plasmid transfer, ...) or highly simplified, summarizing all of these steps into a single rate. The desired complexity will depend on the specific question and context that is studied. More complex models usually bring the difficulty that realistic parameters are not well known for all of the steps and can be hard to determine experimentally. Furthermore, increased complexity may only result in marginal differences in outcomes, while reducing the traceability and interpretability of the model. Therefore, it is crucial to determine the key biological processes relevant to the research question for inclusion in the model. On the other hand, models that are too simple could fail to describe the dynamics seen in nature (see also section \ref{sec:notes}). 

Most mathematical models used for microbial population dynamics are ordinary differential equation (ODE) models. They describe how populations behave over time, for example, how plasmid donor and recipient cell populations grow and exchange plasmids. These models make the crucial assumption that all cells in a particular class/population behave in the same way. ODE models capture the complex dynamics of individual cells through average rates and the overall population size.
Differential equations can be computationally solved in a deterministic or stochastic manner. We are going to focus on deterministic models, where each simulation of the model (with the same parameters and initial conditions) results in the same outcome. These models are easy to work with as they describe average population behavior. However, even in clonal bacterial populations not all cells behave entirely the same and this type of heterogeneity is not captured by deterministic models. Stochastic models can incorporate randomness and variation by introducing noise terms or working with probability distributions instead of fixed parameters. They are more complex to analyze but capture the behavior of small populations and random processes, like the fixation of mutations, much better than deterministic models.  

Agent-based models (ABMs) are a popular alternative for understanding microbial population behavior, especially in environments with spatial structure (i.e. environments that are not well-mixed). ABMs focus on the behavior of individual agents, which allows them to capture spatial and temporal heterogeneity, leading to complex emergent population dynamics. ABMs have several drawbacks, including being computationally intensive, challenging to parameterize, and not straightforward to analyze in terms of linking parameter changes to population level effects. ABMs can  be implemented in both a deterministic or a stochastic manner but the latter is more commonly used. 

Overall, there is not necessarily one `right' mathematical model to describe a biological system. There are many possible ways to construct a model, each with its own advantages and disadvantages. The choice of model type and complexity depends on the use case and is guided by the research question, as well as considerations of computational resources, ease of implementation, analysis, and interpretability. However, in many applications, ODEs are a powerful first approach to better understand the dynamics of a biological system. In the following, we develop an ODE model of plasmid dynamics. This choice of model type is motivated by the aim to understand basic qualitative behavior of plasmid dynamics without using overly complex approaches. Accordingly, we represent complex biological processes, such as conjugation and segregation loss, using simplified parameters, as described in section \ref{sec:one_plasmid_model}. We build an intuition for modeling plasmid dynamics by first developing a model that describes the dynamics of one plasmid (section \ref{sec:one_plasmid_model}) and then incorporating a second plasmid into the mathematical model to follow plasmid co-infection dynamics (section \ref{sec:two_plasmid_model}).

\section*{Infection dynamics of one conjugative plasmids}
\subsection{Developing an ODE model for single plasmid dynamics } \label{sec:one_plasmid_model}

\subsubsection{Drawing a model schematic}
\begin{figure}[ht!]
\centering
\includegraphics[width=0.7\linewidth]{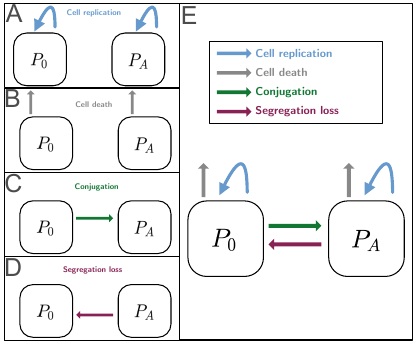}
\caption{\textbf{Model schematic illustrating the biological processes of a bacterial population with one conjugative plasmid.} $P_0$ denotes plasmid-free cells and $P_A$ cells that carry plasmid variant A. Sub-panels illustrate how to depict cell replication \textbf{A}, cell death \textbf{B}, conjugation \textbf{C}, and segregation loss \textbf{D}, while \textbf{E} summarizes all processes in one model schematic.} \label{fig:schematic_one_plasmid}
\end{figure}

We want to develop a mathematical model that tracks the dynamics of a bacterial population through which a conjugative plasmid is spreading. To do so we use ordinary differential equations, which track changes in bacterial cell numbers over time based on bacteria- and plasmid-associated processes (e.g. growth and conjugation). As we want to follow the dynamics of one bacterial species and one conjugative plasmid, our model tracks the changes of two types of bacterial cells: plasmid-free cells $P_0$ and plasmid-carrying cells $P_A$, which are infected with plasmid $A$. Before developing the differential equations $\frac{dP_0}{dt}$ and $\frac{dP_A}{dt}$ describing the change of $P_0$ and $P_A$ over time, it can help to depict the processes that affect these subpopulations in a graphical summary, the so-called model schematic. In a model schematic, the considered populations are represented by boxes, and the biological processes that increase or decrease the size of these populations are indicated with arrows (Fig. \ref{fig:schematic_one_plasmid}A-D). The type of model we are developing here is also called a compartmental model, as it assigns individual cells to populations — referred to as compartments, which are depicted as boxes in the model schematic — and tracks the dynamics of these compartments as a whole rather than the dynamics of each individual cell. For our one plasmid model, we consider the following biological processes that affect the densities of our populations $P_0$ and $P_A$: \\

\textbf{Cell replication} The densities of both populations, $P_0$ and $P_A$ increase due to cell replication. In model schematics, cell replication is often represented as loops that emerge from and point towards the corresponding population box, since the amount of replication (and hence the increase in cell numbers) depends on the population itself (Fig. \ref{fig:schematic_one_plasmid}A).

\textbf{Cell death} Bacterial cells from both populations can also die, leading to a decrease in the respective cell numbers. Death is typically depicted by an arrow pointing away from the corresponding population (Fig. \ref{fig:schematic_one_plasmid}B).

\textbf{Conjugation}
As a result of conjugation a plasmid-free cell transitions to a plasmid-carrying cell, which is indicated by an arrow going from the $P_0$ to the $P_A$ population (Fig. \ref{fig:schematic_one_plasmid}C).

\textbf{Segregation loss}
If a plasmid-carrying cell loses a plasmid, it transitions from the population $P_A$ to the plasmid-free population $P_0$, which is indicated by an arrow from $P_A$ to $P_0$ (Fig. \ref{fig:schematic_one_plasmid}D).

Putting all the individual processes together, we get the complete schematic of our model (Fig. \ref{fig:schematic_one_plasmid}E). \\

\subsubsection{Translating a model schematic into differential equations} \label{ch:develop_one_plasmid_model}

With the help of the model schematic, we can develop the differential equations $\frac{dP_0}{dt}$ and $\frac{dP_A}{dt}$ that describe the change in our populations over time. For this, we need to translate the arrows in the model schematic (figure \ref{fig:schematic_one_plasmid}E) into mathematical expressions. One arrow does not necessarily translate into one unique mathematical representation, but each mathematical expression is based on assumptions and simplifications of biological processes, which should be clearly stated in the model description. In the following, we will first explain the translation from schematic into mathematical expressions and then go into further detail about the underlying model assumptions and simplifications in section \ref{sec:simplifcations}.

\textbf{Cell replication}
Replication can generally be described by a replication rate, which is determined by environmental factors like the nutrient environment. In our model, we do not track the nutrients bacteria grow on explicitly, but rather implicitly: we assume that bacteria follow logistic growth, where cell replication decreases with increasing bacterial density until the carrying capacity $K$ is reached. The growth of our plasmid free population therefore can be described as
{\small
\begin{align*} 
	\begin{split}
		\frac{dP_0}{dt} =& P_0\left[\textcolor{replication}{\rho_0(1-\frac{T}{K})}\right],
	\end{split}
\end{align*}
}
where $\rho_0$ is the maximal replication rate of the plasmid-free population $P_0$ and $T= P_0 +P_A$ the total cell density. The growth of plasmid-carrying cells can be described equivalently, but with a different maximal replication rate $\rho_A$, which represents the fitness effect plasmid carriage has on the bacterial host. Often, plasmid maximal replication rates are represented in an equivalent, but more explicit, form: $\rho_A= (1-c) \rho_0 $, where the factor $c$ represents a plasmid cost if $ 0\leq c< 1$ and a plasmid benefit if $c<0$.  \\

\textbf{Cell death}
We model cell death with a death rate $\gamma_i$, with $i$ representing $\gamma_0$ or $\gamma_A$. Here, we again assume that plasmid-free and plasmid-carrying cells can have different death rates due to plasmid carriage, reflecting a benefit (e.g. antibiotic resistance) or cost. We assume that death does not depend on the total cell density $T$, but only on the corresponding cell type density $P_0$ or $P_A$. Cell death can therefore be described as
{\small
\begin{align*} 
	\begin{split}
		\frac{dP_i}{dt} =& P_i\textcolor{death}{ \gamma_i}  ,
	\end{split}
\end{align*}
}
where $i$ can refer to $0$ or $A$. \\

\textbf{Plasmid conjugation}
Conjugation requires physical interaction between plasmid-carrying and plasmid-free cells. We assume a well-mixed bacterial population, where conjugation events for plasmid $A$ occur in a density-dependent manner with conjugation rate $\beta_A$. The decrease of the plasmid-free population through conjugation can be described as 
{\small
\begin{align*} 
	\begin{split}
		\frac{dP_0}{dt} =& -\textcolor{conjugation}{\beta_A} P_0 P_A.
	\end{split}
\end{align*}
}
While this term is subtracted in the differential equation for plasmid-free cells, the same term is added in $\frac{dP_A}{dt}$ as new plasmid-carriers are created, i.e.
{\small
\begin{align*} 
	\begin{split}
		\frac{dP_A}{dt} =& \textcolor{conjugation}{\beta_A} P_0 P_A .
	\end{split}
\end{align*}
}
In general, mathematical terms for processes that cause transitions between cell types $P_0$ and $P_A$, must be subtracted in one differential equation and added in the other.\\

\textbf{Segregation loss}
In our model, we link segregation loss to cell replication: we assume that during replication there is a probability $s_A$ that the plasmid is not inherited by both daughter cells, resulting in one plasmid-carrying and one plasmid-free daughter cell. The increase of plasmid-free cells through plasmid segregation loss can be described as
{\small
\begin{align*} 
	\begin{split}
		\frac{dP_0}{dt} =& \textcolor{loss} { (1-\frac{T}{K})\left[\rho_A s_A P_A\right]},
	\end{split}
\end{align*}
}
and the decrease of plasmid-carrying cells can be described as
{\small
\begin{align*} 
	\begin{split}
		\frac{dP_A}{dt} = - & \textcolor{loss} { (1-\frac{T}{K})\left[\rho_A s_A P_A\right]}.
	\end{split}
\end{align*}
}\\
Note the same logistic form as in the cell replication process here, as segregation loss is coupled to effective growth rates and assumed to be zero at carrying capacity $K$.

\textbf{An ODE model describing the dynamics of a single plasmid}
By combining the mathematical expressions of the individual biological processes, we obtain the following two differential equations to describe the dynamics of one bacterial species with a single conjugative plasmid:

{\small
\begin{align}\label{eq:single_plasmid_model}
	\begin{split}
		\frac{dP_0}{dt} =& P_0\left[\textcolor{replication}{\rho_0(1-\frac{T}{K})}  \textcolor{death}{- \gamma_0}  \textcolor{conjugation}{- \beta_A}P_A  \right] \textcolor{loss} {+ (1-\frac{T}{K})\left[\rho_A s_AP_A\right]}\\
		\frac{dP_A}{dt}=& P_A\left[\textcolor{replication}{\rho_A(1-s_A)(1-\frac{T}{K})} - \textcolor{death}{\gamma_A }  + \textcolor{conjugation}{\beta_A} P_0\right] \\
	\end{split} 
\end{align}
}

\FloatBarrier
\subsection{Simplifications and assumptions } \label{sec:simplifcations} 

Several assumptions were made to arrive at the equations in the previous section. Here we spell out these simplifications and assumptions explicitly, and give some guidance on alternative assumptions that could be made. 

\subsubsection*{Cell replication}
In the model, we assume that the different populations follow logistic growth. This means that the per capita growth rate of cells is highest at low density, and slows down as the population reaches carrying capacity. This is a common assumption to describe what might happen as an increasingly large population depletes the resources that are available in the environment. Note that nutrients can also be modeled explicitly if it is crucial to capture bacterial metabolism in the system under investigation. 

In simple logistic growth models, the carrying capacity is simultaneously the maximal size that a population can be stably maintained at. However, since we include a death term in our population equations, the maximal population size is actually a little bit lower. We can see this by finding the steady state for the plasmid-free population ($P_0$) in the absence of the plasmid-carrying population ($P_A=0$). At the steady state, the cell density does not change over time, i.e., equation \ref{eq:single_plasmid_model} 
equals zero:
\begin{align*}
\frac{dP_{0}}{dt} &= 0 \\
&\Leftrightarrow \frac{dP_{0}}{dt} = P_0 (\rho_0(1-\frac{P_0}{K})-\gamma_0) = 0 \\
&\Leftrightarrow P^{*}_{0} = 0 \hspace{1cm}\text{or} \hspace{1cm} P^{*}_{0} = (1-\frac{\gamma_0}{\rho_0})K
\end{align*}

Whereas without a death term, the steady state population sizes would be $P^{*}_{0} = 0$ and $P^{*}_{0} = K$.

We further assume that both plasmid-carrying and plasmid-free populations are counted together (\emph{T}) to describe the growth rate slowdown near the carrying capacity. This reflects an assumption that both populations consume the same resources. If a plasmid variant carries genes enabling its host to use additional resources, a way to describe the effects of reduced resource competition would be to substitute the term $K$ by $K_{i}$. The ability to consume new resources may also bring more complex growth behaviors with it, which are not captured in the parameter of carrying capacity alone. In this case also other aspects of the logistic growth may need to be modified.

We describe fitness costs of the plasmid in a multiplicative manner ($\rho_A= (1-c) \rho_0 $). Translated into biological terms, this means that the presence of the plasmid reduces the growth rate by a certain percentage rather than an absolute amount. 

\subsubsection*{Cell death}
We model cell death separately, and independent of the carrying capacity, to ensure continuous population turnover at any density. 

Taken together with the growth term, this means that the net growth of a population (birth - death) is made up of both density-dependent and -independent terms. Without the density-independent component of the cell death term, cells would not replicate or die at carrying capacity. In that case, the effect of plasmid carriage on host cell fitness would no longer matter once carrying capacity is reached, and there would be no plasmid loss as segregation is linked to replication. 

This has important implications for any evolutionary predictions derived from the model. For example, if a plasmid affects host growth (e.g. due to antibiotic resistance carriage), not including a density-independent growth term would allow plasmid-free and plasmid-carrying cells to coexist once carrying capacity is reached even if one cell type may out-compete the other in the presence of constant population turnover.

\subsubsection*{Conjugation}
Conjugation is assumed to follow mass action kinetics~\cite{Levin1979TheModel}, at a constant rate and proportional to the densities of both plasmid-free and plasmid-carrying populations. This is a simplification, which was empirically demonstrated for F plasmids in well-mixed liquid culture~\cite{Levin1979TheModel}. 

In reality, the conjugation parameter $\beta$ is a compound parameter summarizing the rate at which cells encounter each other (which may depend on environmental parameters including mixing speed) and the per-encounter probability of successful plasmid transfer~\cite{Zbinden2025}.

In particular, this is a poor assumption for cells that grow in structured environments such as biofilms. There, donor cells cannot be assumed to encounter all recipient cells with equal probability. It may, however, be a sufficiently good approximation when cells are grown together on filters for a short amount of time and at sufficiently high density~\cite{Zhong2012}.

It also ignores the fact that the conjugation rate is not necessarily constant in time. Mobile genetic elements, like plasmids, are often dependent on the machinery of their host cell for replication and conjugation. This means that conjugation rate could change over time, in dependence of metabolic changes in their host cell.

\subsection{Implementation} 

Now that we have established the model equations, we can use them to explore plasmid dynamics in scenarios of interest. To do this, we will implement the model using a programming language such as R (see section \ref{full_R_code}), and run simulations with specialized packages like deSolve (see section \ref{sec:materials}). To obtain meaningful results, we need to assign values to the model parameters and set initial population numbers based on the scenario we want to study. 

\subsubsection{Parametrization}
Depending on the research question and the empirical data available, the parameters values can be derived from laboratory experiments or drawn from existing literature. In cases where literature values are not available for the system under study, extrapolation from similar systems may be necessary. However, since the primary aim of theoretical studies is often to explore a general concept or theory in a qualitative manner, absolute values might be less relevant than the relative magnitudes of parameters (e.g. conjugation rate versus growth rate). Additionally, these studies often involve sensitivity analysis, where a broad range of reasonable parameter values are tested to assess how sensitive the results are to variations in the parameters. This not only helps determine whether the theory is broadly applicable, but also highlights which parameters significantly influence the system’s dynamics.
For theoretical studies, parameters and the total population size can be rescaled to fall between 0 and 1 to facilitate interpretation of results from a wide range of simulations where parameters were based on different experimental setups and biological systems. However, care must be taken to preserve the relative magnitudes and units (e.g., time scales) during the scaling process.   

The relative scale of different parameters can also shape the model’s structure. For instance, if certain processes occur much faster than others, their detailed dynamics may be neglected and instead approximated by constant values, a technique known as the quasi-steady-state approximation. In contrast, if certain processes involve delays (e.g. the time it takes to transfer a plasmid between cells), the assumption of a continuous rate may not be appropriate. Whether to explicitly incorporate delays depends on the specific context and time scale of the model.

Fitting parameter values to experimental data typically aims to describe and explain the behavior of a specific system under certain conditions (for a reference on data fitting see section \ref{sec:notes}). Data fitting can be particularly powerful when used to generate realistic parameter values, especially if parameters derived from one dataset can predict outcomes in another. The success of such predictions provides insight into how well the model generalizes beyond the initial dataset. However, for models with a large number of parameters, obtaining meaningful fits can be challenging, particularly when dealing with noisy data. Furthermore, if the model is not able to reproduce qualitative behavior of experimental results, it suggests that the mathematical model might be missing essential processes and that our reasoning to explain biological reality is incomplete.

\subsubsection{Initial conditions}
Initial conditions define the starting point of model simulations and must be selected carefully, as they can play a critical role in determining the long-term behavior of the system. In other cases, the initial conditions primarily influence how quickly a qualitative result is reached (illustrated in Fig. \ref{fig:segregation}). When the model is fitted to empirical data, estimates for the initial conditions follow from knowledge about the experimental setup. However, for theoretical scenarios, initial conditions are often guided by the specific research question. For example, when studying plasmid invasion, it is reasonable to start with $P_A$ at a very low percentage of the population (<1\%). 
As with parameter values, it is often beneficial to run simulations with a range of initial conditions to assess the sensitivity of the results to different starting points. This helps ensure that the model's conclusions are robust and not overly dependent on specific assumptions.

\subsubsection{Analyzing the simulation results}
\begin{figure}[ht!]
\centering
\includegraphics[width=1\linewidth]{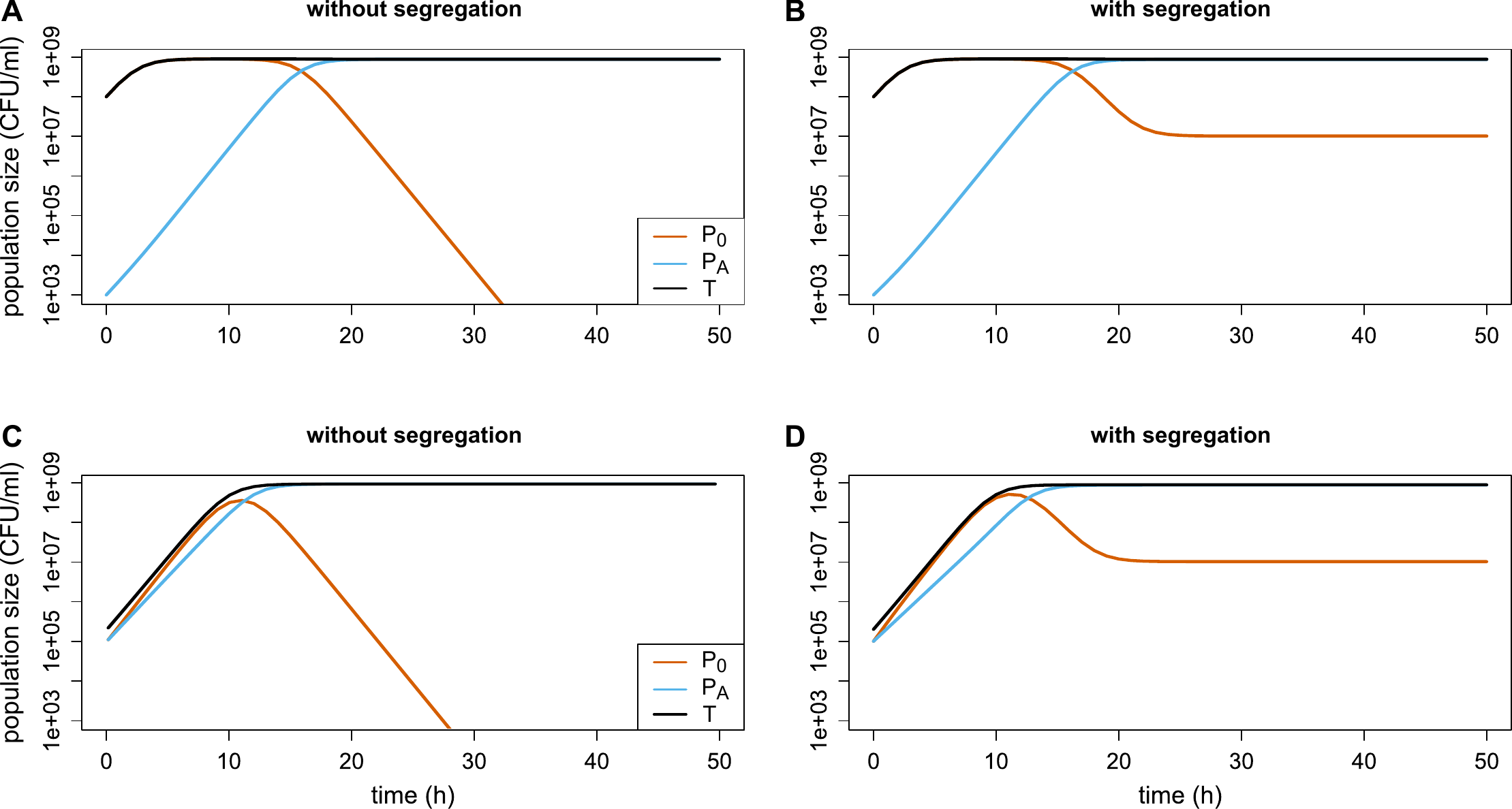}
\caption{\textbf{Simulation results for single plasmid dynamics.} Deterministic simulations of a single plasmid spreading in a population without (A,C) and with (B,D) plasmid segregation rate, when the plasmid is invading from rare (A,B) or competing at equal densities (C,D). Plasmid-free ($P_0$) cells are shown in orange, plasmid-carrying cells in blue ($P_A$) and the total number of cells in black. Simulations were run over 50h for $\rho_0=1 \mathrm{h}^{-1}, \rho_A = 0.85 \mathrm{h}^{-1}, \gamma_0 = \gamma_A =0.1 \mathrm{h}^{-1}, K=10^9 \frac{\mathrm{CFU}}{\mathrm{ml}}, \beta_A = 10^{-9} \frac{\mathrm{ml}}{\mathrm{CFU \cdot h}}$, in A and C:  $s_A = 0 $, in C and D: $s_A=0.1 $, in A and B: $P_{0}(0) = 10^8\frac{\mathrm{CFU}}{\mathrm{ml}}, P_{A}(0) = 10^3\frac{\mathrm{CFU}}{\mathrm{ml}}$, in C and D: $P_{0}(0) = 10^5\frac{\mathrm{CFU}}{\mathrm{ml}}, P_{A}(0) = 10^5\frac{\mathrm{CFU}}{\mathrm{ml}}$.} \label{fig:segregation}
\end{figure}

We now present examples of simulation runs using the model equations (\ref{eq:single_plasmid_model}) developed in section \ref{ch:develop_one_plasmid_model}. In these simulations, we track the dynamics of two bacterial populations: plasmid-free cells ($P_0$) and plasmid-carrying cells ($P_A$). The results are visualized by plotting the population dynamics over time (Fig. \ref{fig:segregation}).

To begin, we explore the spread of a costly plasmid (with a 15\% reduction in growth rate) in a population initially consisting largely of plasmid-free cells. We observe that the plasmid successfully spreads and dominates the population (Fig. \ref{fig:segregation}A). This is due to the ability of the plasmid to replicate not only via vertical transmission but also through horizontal transmission, as conjugation rates are sufficiently high to enable the spread of the costly plasmid. 

This effect is particularly strong, because we did not account for plasmid loss ($s_A=0$) in this first set of simulations. When we introduce segregation loss of plasmids into the simulations ($s_A=0.1$), we observe that the plasmid-carrying population still dominates, though it now coexists with a subpopulation of plasmid-free cells (Fig. \ref{fig:segregation}B). Further increasing the segregation rate would increasingly favor the plasmid-free population over the plasmid-carriers. 

To understand the impact of initial conditions in this scenario, we investigate a second ecological scenario, where the plasmid does not invade from rare but is competing at equal densities with plasmid-free cells, i.e. we vary our initial conditions from $P_{0}(0) = 10^8, P_{A}(0) = 10^3$ (Fig. \ref{fig:segregation}C) to $P_{0}(0) = 10^5, P_{A}(0) = 10^5$ (Fig. \ref{fig:segregation}D). We find that the plasmid takes over the population more quickly in the competition scenario. However, the final equilibrium states are the same in both cases, indicating that the initial conditions do not affect the long-term dynamics under these conditions (Fig. \ref{fig:segregation}). 

\FloatBarrier
\section*{Co-infection dynamics of two conjugative plasmids}
To capture the dynamics of two conjugative plasmids in a mathematical model, we extend the one-plasmid model developed in section \ref{sec:one_plasmid_model}. In the following, we show how part of the two-plasmid model can be derived directly from the one-plasmid model by adding analogous equations for the second plasmid, and we emphasize how potential interactions between the plasmids add additional complexity to the two-plasmid model.

\subsection{Relevant populations and processes – drawing a model schematic} \label{sec:two_plasmid_model}


\begin{figure}[ht!]
\centering
\includegraphics[width=0.8\linewidth]{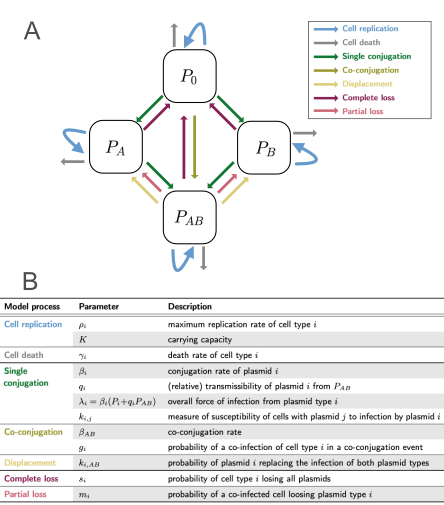}
\caption{\textbf{Visualization of the modeled plasmid co-infection processes and the corresponding parameters.} \textbf{A.} Schematic diagram of the co-infection model given by equations \ref{co-infection model}. $P_0$ denotes plasmid-free cells, $P_A$ and $P_B$ are bacterial cells infected with plasmid variant $A$ or $B$, respectively, and $P_{AB}$ are cells co-infected with $A$ and $B$. Arrows indicate the transition of cells between states. \textbf{B.} Co-infection processes incorporated in the model, listed with their associated parameters and parameter descriptions. This figure is taken from our previous work \textit{Igler et al. 2022} \cite{igler2022}, which is published under the CC BY 4.0 license.} \label{fig:schematic}
\end{figure}

To obtain the plasmid co-infection model with two plasmids from the one-plasmid model developed in section \ref{sec:one_plasmid_model}, we need to increase the number of tracked cell types to four.
Adding another plasmid variant $B$ introduces the compartment of cells infected with plasmid type $B$ ($P_B$) and cells infected with both plasmid variants ($P_{AB}$). The occurrence of cells co-infected with both plasmid variants, depends on the biological properties of the individual plasmids and how these properties are incorporated into the model assumptions. Here, we develop a general model of plasmid co-infection that provides a comprehensive summary of potential plasmid interactions and can later be tailored to specific plasmids of interest by choosing model parameters accordingly. In the following, we formalize the extension of the one-plasmid model from section \ref{sec:one_plasmid_model} to a plasmid co-infection model with two plasmids, while simultaneously extending the model schematic of the one-plasmid model (Fig. \ref{fig:schematic_one_plasmid}E) to the plasmid co-infection model schematic (Fig. \ref{fig:schematic}A).

\textbf{Cell replication} 
As in the one-plasmid model schematic the bacterial growth for each of the four bacterial cell types is represented as a self-loop (Fig. \ref{fig:schematic}A). Mathematically, bacterial growth can be described as previously by $\rho_i(1-\frac{T}{K})$, where $\rho_i$ represents the maximum replication rate of cell type $i$ ($i$=$0$, $A$, $B$ or $AB$), $T$ the total cell density ($T = P_0 + P_A + P_B + P_{AB}$) and $K$ the carrying capacity of all cell types. 

\textbf{Cell death} In the co-infection model schematic, cell death is depicted, as before, by arrows pointing away from each individual cell type \ref{fig:schematic}A). The death term for each cell type is given by $\gamma_i P_i$, where $\gamma_i$ is the death rate of cell type $i$.

\textbf{Plasmid conjugation} 
Generally, as in the one-plasmid model, conjugation events are depicted in the model schematic as arrows pointing from the recipient cell type before conjugation to the cell type after conjugation. Conjugation from a singly infected cell ($P_A$ and $P_B)$ to plasmid-free cells $P_0$ is analogous to the one-plasmid model, that is $\beta_i P_i P_0$ with $i=A$ or $B$. However, conjugation involving co-infected donor cells $P_{AB}$ or already infected recipient cells is not as straightforward. These conjugation events require careful consideration of plasmid interactions and model structure to avoid unintended biases, which favor plasmid co-infection \cite{igler2022}. Here, we conceptualize the potentially complex processes associated with plasmid co-infection into a general mathematical formulation, with a more detailed explanation about biological background and assumptions given in section \ref{sec:assumptions_two_plasmid_model} or also in \textit{Igler et al. 2022} \cite{igler2022}.


\underline{\textit{Co-infected cells $P_{AB}$ as donors:}}
We assume that during a conjugation event a co-infected cell can either transmit a single plasmid type ($A$ or $B$) to a plasmid-free cell or both plasmids ($A$ and $B$) at the same time. Transmission of both plasmids together occurs with the conjugation rate $\beta_{AB}$. The single conjugation rate $\beta_i$ ($i=A$ or $B$), previously used to describe the transmission of plasmids from a singly infected cell, may be altered when conjugation of a single plasmid occurs from a co-infected cell due to plasmid interactions caused e.g. by fertility inhibition systems (for more detail see \textit{Igler et al. 2022} \cite{igler2022}). Hence, the single conjugation rate from a plasmid co-infected cell in our model is given by $q_i \beta_i P_0 P_{AB}$, where the parameter $q_i$ reflects the relative transmissibility of plasmid $i=A$ or $B$ from a co-infected cell. Overall, the decrease in density of plasmid-free cells due to conjugation initiated by co-infected cells is given by

{\small
\begin{align*} 
	\begin{split}
		\frac{dP_0}{dt} =& -\left(\textcolor{conjugation}{q_A \beta_A +q_B \beta_B } \textcolor{co-conjugation}{+\beta_{AB}} \right) P_0 P_{AB},
	\end{split}
\end{align*}
}
which leads to an increase in the following populations
{\small
\begin{align*} 
	\begin{split}
		\frac{dP_A}{dt} = \textcolor{conjugation}{q_A \beta_A}P_0 P_{AB},\\ 
        \frac{dP_B}{dt}= \textcolor{conjugation}{q_B \beta_B } P_0 P_{AB}, \\
        \frac{dP_{AB}}{dt}=\textcolor{co-conjugation}{\beta_{AB}} P_0 P_{AB} .
	\end{split}
\end{align*}
}

The fact that a single plasmid can be acquired both from singly infected cells and from co-infected cells can be summarized as the `force of infection', a concept from epidemiology. The force of infection is the rate at which susceptible cells acquire a plasmid in the population. The force of infection of plasmid $i$ ($i=A$ or $B$) is given by
{\small
\begin{align*} 
	\begin{split}
		\lambda_i =& \beta_i \left(P_i+q_i P_{AB} \right).
	\end{split}
\end{align*}
}
The force of infection is not only useful as a concept to describe an overall transmission-rate of a single plasmid type, but it also makes the equations of our plasmid co-infection model more compact. For example, the density decrease of the plasmid-free population due to conjugation can be summarized in the following compact form using the force of infection
{\small
\begin{align*} 
	\begin{split}
	\frac{dP_0}{dt} =& -P_0\left[ \textcolor{conjugation}{ \lambda_A +\lambda_B}  \textcolor{co-conjugation}{- \beta_{AB}P_{AB}}\right],
	\end{split}
\end{align*}
}
instead of the long version
{\small
\begin{align*} 
	\begin{split}
	\frac{dP_0}{dt} =& -P_0\left[ \textcolor{conjugation}{ \beta_A \left(P_A+q_A P_{AB} \right) + \beta_B \left(P_B+q_B P_{AB} \right) }  \textcolor{co-conjugation}{+ \beta_{AB}P_{AB}}\right].
	\end{split}
\end{align*}
}


\underline{\textit{Plasmid infected cells as recipients:}} In contrast to the one-plasmid model, the plasmid co-infection model also needs to consider conjugation events where the recipient already carries a plasmid, leading to possible plasmid interactions at conjugation. 

\textit{Singly infected cells ($P_A$ or $P_B$) as recipients:}
We assume that if a cell is already infected with one plasmid $j$ ($j=A$ or $B$) then the susceptibility to infection with a further plasmid type $i$ ($i=B$ or $A$) is altered compared to plasmid-free cells. This altered susceptibility can stem, for example, from exclusion systems (more details in \textit{Igler et al. 2022} \cite{igler2022}) and is summarized in the parameter $k_{i,j}$. Hence, the decrease in cell density of cells carrying plasmid type $A$ by receiving plasmid type $B$ via single conjugation from the donor cells $P_B$ and $P_{AB}$ can be described as
{\small
\begin{align*} 
	\begin{split}
	\frac{dP_A}{dt} =& \textcolor{conjugation}{ -k_{B,A}\lambda_B P_A }.
	\end{split}
\end{align*}
}
So far, we only considered the single transfer of plasmid type $B$, however as introduced above, we assume that co-infected cells $P_{AB}$ can co-transfer both plasmid types with rate $\beta_{AB}$. We assume that co-conjugation to an already infected cell $P_A$ (or $P_B$) leads to co-infection with probability $g_A$ (or $g_B$). Therefore, the total decrease in cell density of $P_{A}$ through conjugation is described by 
{\small
\begin{align*} 
	\begin{split}
	\frac{dP_A}{dt} =& - k_{B,A}P_A(\textcolor{conjugation}{\lambda_B } \textcolor{co-conjugation}{+g_A\beta_{AB}P_{AB}}).
	\end{split}
\end{align*}
}
Note that the same conjugation events lead to a cell density increase of co-infected cells $P_{AB}$ and therefore this term needs to be added to the differential equation of $P_{AB}$. The conjugation terms describing the transmission of plasmid type $A$ to $P_B$ are analogous.

\textit{Co-infected cells ($P_{AB}$) as recipients:}
We also need to consider plasmid conjugation to co-infected cells. Conjugation of a plasmid variant into co-infected cells can lead to \textit{displacement} of the other plasmid variant and hence a transition of co-infected cells to singly infected cells. This assumes that the cell-internal processes leading to loss of the other plasmid type occur much faster than the other population-level processes such as growth and conjugation (more details see section \ref{sec:assumptions_two_plasmid_model} and \textit{Igler et al. 2022} \cite{igler2022}). The decrease in density of co-infected cells through displacement as a result of conjugation can be described as 
{\small
\begin{align*} 
	\begin{split}
		\frac{dP_{AB}}{dt} =& \textcolor{displacement}{-\left( k_{A,AB}\lambda_A - k_{B,AB}\lambda_B\right) P_{AB}}, \\
	\end{split}
\end{align*}
}
where $k_{i,AB}$ denotes the probability of plasmid $i=A$ (or $B$) displacing plasmid $j=B$ (or $A$) from a co-infected cell due to conjugation. The displacement leads to an increase of singly infected cells
{\small
\begin{align*} 
	\begin{split}
		\frac{dP_A}{dt} = \textcolor{displacement}{k_{A,AB}\lambda_A P_{AB} } \text{, and } \frac{dP_B}{dt}= \textcolor{displacement}{k_{B,AB}\lambda_BP_{AB}}      .
	\end{split}
\end{align*}
}

\textbf{Plasmid segregation loss} 
As in the one-plasmid model, segregation loss is represented in the model schematic as an arrow pointing from the cell type that loses the plasmid to the cell type resulting from plasmid loss. In the co-infection model we distinguish between complete segregation loss and partial segregation loss, as plasmid loss for co-infected cells does not necessarily result in plasmid-free cell types:

 \textit{Complete loss:} At cell replication plasmid carriage can be completely lost with probability $s_i$ ($i=A,B \text{, or } AB$). The mathematical equations for complete loss are analogous to the one-plasmid model.

\textit{Partial loss:} 
Co-infected cells lose one of the plasmids at cell replication with probability $m_i$ ($i=A$ or $B$). As we assume that segregation is linked to cell replication, the number of cells that lose the plasmid cannot exceed the number of cells produced at replication, hence there is the constraint $m_A + m_B \leq 1$. To observe single loss of a co-infected cell, the cell has to lose one plasmid, but not both, therefore this event is observed with probability $m_i \left( 1-s_{AB}\right)$ ($i=A$ or $B$). Hence, the reduced growth of co-infected cells due to partial plasmid loss can be described by
{\small
\begin{align*} 
	\begin{split}
		\frac{dP_{AB}}{dt} =& \textcolor{replication}{\rho_{AB}(1-s_{AB})(-m_A-m_B)(1-\frac{T}{K}) P_{AB}}. \\
	\end{split}
\end{align*}
}
Depending on the mechanism causing plasmid loss in co-infected cells, partial and complete loss may not be independent - which can be captured by additional constraints on $m_i$ and $s_{AB}$. However, similar to the constraint for partial plasmid loss, the total probability of plasmid loss is always restricted by $s_{AB}+m_A+m_B\leq 1$, regardless of whether partial and complete loss is independent or not. Dependencies of partial and complete loss, such as the loss of one plasmid facilitating the loss of the other, can be represented by the additional constraint $s_{AB} \geq m_A m_B$. In contrast, if losing one plasmid makes it less likely to lose the other, then $s_{AB} \leq m_A m_B$.
Total plasmid loss from co-infected cells is given by
{\small
\begin{align*} 
	\begin{split}
		\frac{dP_{AB}}{dt} =& \textcolor{replication}{\rho_{AB}(1-s_{AB})(1-m_A-m_B)(1-\frac{T}{K}) P_{AB}}. \\
	\end{split}
\end{align*}
}
Note that the corresponding loss terms need to be added in the differential equations of $P_0$, $P_A$, and $P_B$, i.e. the cell types that gain in density due to segregation loss of co-infected cells. 

Overall, the plasmid co-infection model are summarized graphically with the model schematic depicted in Fig. \ref{fig:schematic} A and the parameters associated to the individual model processes in Fig. \ref{fig:schematic} B.


The plasmid co-infection model schematic (Fig. \ref{fig:schematic} A) can be mathematically formalized by the following differential equations:
{\small
\begin{align} \label{co-infection model}
	\begin{split}
		\frac{dP_0}{dt} =& P_0\left[\textcolor{replication}{\rho_0(1-\frac{T}{K})}  \textcolor{death}{- \gamma_0}  \textcolor{conjugation}{- \lambda_A - \lambda_B}  \textcolor{co-conjugation}{- \beta_{AB}P_{AB}}\right] \textcolor{loss} {+ (1-\frac{T}{K})\left[\rho_A s_AP_A + \rho_B s_B P_B + \rho_{AB} s_{AB} P_{AB}\right]}\\
		\frac{dP_A}{dt}=& P_A\left[\textcolor{replication}{\rho_A(1-s_A)(1-\frac{T}{K})} - \textcolor{death}{\gamma_A }  - k_{B,A}(\textcolor{conjugation}{\lambda_B } \textcolor{co-conjugation}{+g_A\beta_{AB}P_{AB}})\right]  + \lambda_A(\textcolor{conjugation}{P_0}\textcolor{displacement}{ + k_{A,AB}P_{AB}}) \\ & \textcolor{partial loss}{+ m_B\rho_{AB}(1-s_{AB})(1-\frac{T}{K})P_{AB}}\\
		\frac{dP_B}{dt} =& P_B\left[\textcolor{replication}{\rho_B(1-s_B)(1-\frac{T}{K})}  \textcolor{death}{- \gamma_B } - k_{A,B}(\textcolor{conjugation}{\lambda_A } \textcolor{co-conjugation}{+g_B\beta_{AB}P_{AB}})\right]  + \lambda_B(\textcolor{conjugation}{P_0}\textcolor{displacement}{ + k_{B,AB}P_{AB}}) \\ &\textcolor{partial loss}{+ m_A\rho_{AB}(1-s_{AB})(1-\frac{T}{K})P_{AB}}\\
		\frac{dP_{AB}}{dt} =& P_{AB}\bigg[ \bigg. \textcolor{replication}{\rho_{AB}(1-s_{AB})(1-m_A-m_B)(1-\frac{T}{K})} \textcolor{co-conjugation}{+ \beta_{AB}(P_0 + g_A k_{B,A} P_A + g_B k_{A,B} P_B)} \\ & \textcolor{death}{- \gamma_{AB}}  \textcolor{displacement}{- k_{A,AB}\lambda_A - k_{B,AB}\lambda_B} \bigg. \bigg] \textcolor{conjugation}{ + k_{B,A}\lambda_BP_A + k_{A,B}\lambda_AP_B}\\
	\end{split}
\end{align}
}

\FloatBarrier
\subsection{Simplifications and assumptions } \label{sec:assumptions_two_plasmid_model}

\subsubsection*{Plasmid copy number}
We do not explicitly model the copy number of a plasmid. Instead, we assume that a plasmid - once introduced into a cell - will quickly equilibrate to the copy number dictated by its replication machinery. If the two co-infecting plasmids $A$ and $B$ are under the same copy number control, this will be reflected in higher segregational loss rates (transitions into the singly infected cell type) from population $P_{AB}$. Similarly, if co-infection increases the copy number of one or both plasmid types, the parameters can be updated to reflect the corresponding biological consequences (e.g. lower segregational loss rates or higher conjugation rates).

\subsubsection*{The nature of co-infection}
Important to note in this model, is that multiple infections with the same plasmid type are equivalent to a single infection with that plasmid as we assume that within the bacterial cytoplasm clonal plasmids will equilibrate to the copy number set by their replication origin. 
This is in contrast to multicellular hosts, where separate infections by infectious agents could occur at multiple sites (e.g. gut, bladder). In such a case the co-infected state is ecologically different from single infection with an infectious disease variant and care must be taken to make sure the resulting epidemiological model reflects this distinction (i.e. is structurally neutral~\cite{Lipsitch2009, igler2022}).

\subsubsection*{Superinfection and displacement}
Our model includes explicit displacement terms ($k_{i, AB}$) to describe the scenario where conjugation of plasmid $A$ or $B$ into a co-infected cell turns that cell into one that is singly infected. This may for instance occur if conjugation leads to a stressed cell state which increases the probability of plasmid segregation for one of the plasmids. However, it is important to note that many, if not most, plasmids are believed to carry superinfection exclusion systems that efficiently exclude related plasmids~\cite{Garcillan-Barcia2008}, reducing co-infection with the same plasmid type significantly. As such the displacement term might be negligibly small for closely related plasmid pairs.

\subsection{Implementation}

Similar to the single plasmid model, we implement our model equations in a programming language (see section \ref{full_R_code}) and run simulations for various scenarios of interest (parameter value combinations). 

\subsubsection{Parametrization}

As we want to study co-infection dynamics, we are most interested in the parameters relating to the relative fitness of the two plasmids, such as costs and conjugation rates, and the direct interactions between them, including infection susceptibility of plasmid carriers. Many of these processes lack precise quantitative data. However, the underlying molecular mechanisms are generally well understood (see the references in \textit{Igler et al. 2022} \cite{igler2022}) and guide how we incorporate the biological effects into the model.
Biological processes can be captured in our developed co-infection model in various ways. For example, the replication or partitioning incompatibility between two plasmids could be represented either as an increased segregation rate for both plasmids or as a reduced susceptibility to co-infection. The choice between these two approaches depends on the assumed mechanism and time scales involved. If plasmid loss occurs gradually, modeling it as an increased segregation rate is more appropriate. On the other hand, if incompatibility prevents the establishment of a second plasmid, reduced susceptibility to co-infection may be a better representation. Both approaches are valid, but it is essential to recognize that these are assumptions, and they will influence the simulation outcomes. Thus, careful consideration is needed when selecting the most appropriate parameter effect for the given biological context.

\subsubsection{Initial conditions}

Depending on the specific research question, co-infection models will typically be initialized with at least one of the two plasmid population at low numbers. This approach helps to examine the invasion dynamics of one or more plasmids into either a plasmid-free population or a population already carrying plasmids. However, we can also envision scenarios where we want to investigate the competition between established plasmids.

\subsubsection{Analyzing the simulation results}

\begin{figure}[ht!]
\centering
\includegraphics[width=1\linewidth]{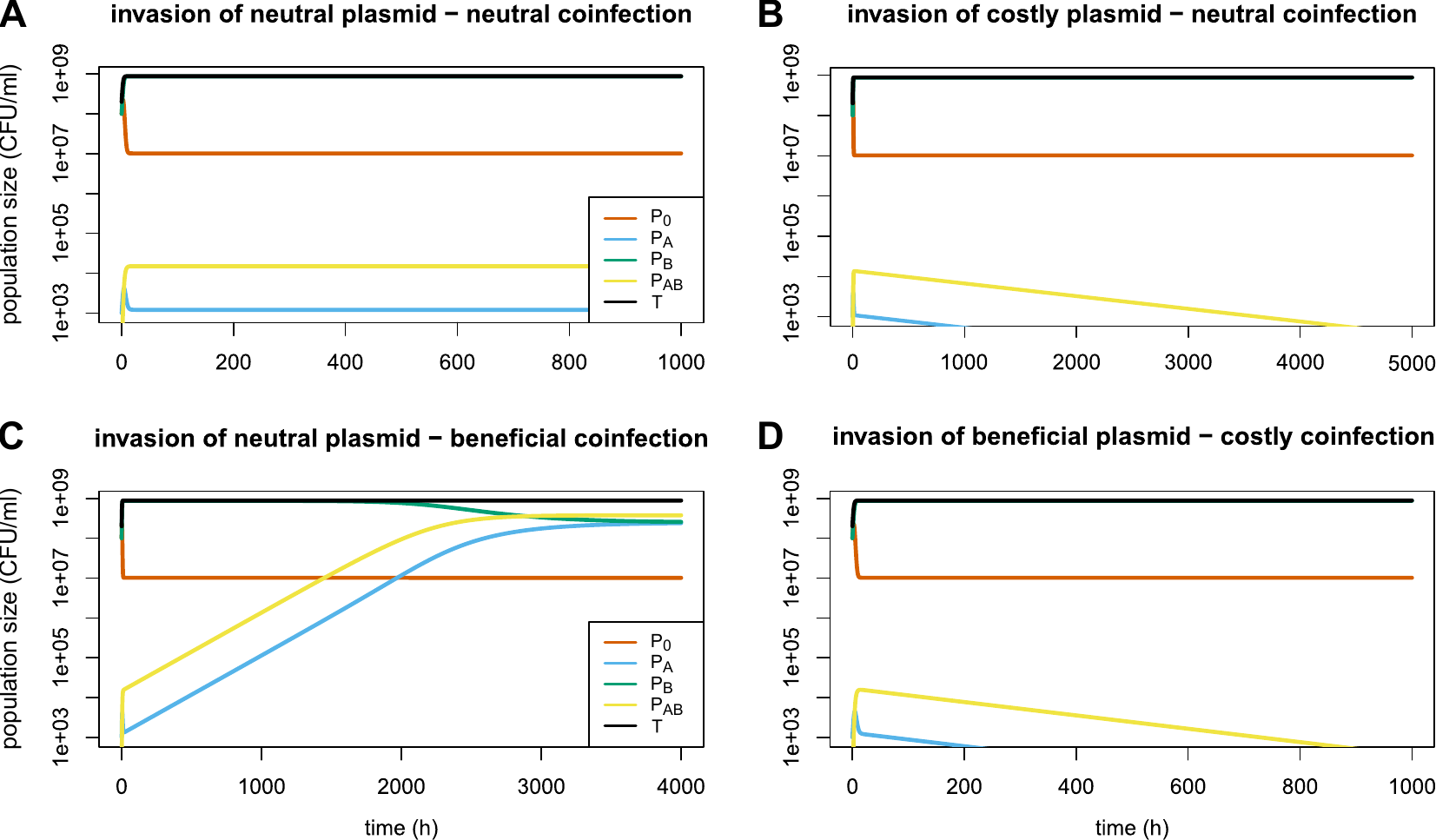}
\caption{\textbf{Simulation results for co-infection dynamics of two plasmids with varying fitness effects of the invading plasmid and the co-infected state.} Deterministic simulations of a plasmid co-infection model where the plasmid A is trying to invade a population in which plasmid B is already prevalent. Dynamics are shown for plasmid A having an equal (A) or lower (B) fitness effect on the host than plasmid B when the co-infection state is equal to the fitness of plasmid B. Invasion is shown for neutral (C) or beneficial (D) plasmid A if the co-infection state is more beneficial (C) or more costly (D) than carrying either A or B. Plasmid-free ($P_0$) cells are shown in orange, cells carrying plasmid A ($P_A$) in blue, cells carrying plasmid B ($P_B$) in green, cells carrying plasmids A and B ($P_{AB}$) in yellow, and the total number of cells in black. Simulations were run for up to 5000h, for for $\rho_0=1 \mathrm{h}^{-1}, \rho_B = 0.85 \mathrm{h}^{-1}, \gamma_0 = \gamma_A = \gamma_B = \gamma_{AB}=0.1 \mathrm{h}^{-1}, K=10^9\frac{\mathrm{CFU}}{\mathrm{ml}}, \beta_A = \beta_B = 10^{-9}\frac{\mathrm{ml}}{\mathrm{CFU \cdot h}}, \beta_{AB} = 0\frac{\mathrm{ml}}{\mathrm{CFU \cdot h}}, s_A = s_B = s_{AB}=0.1 , q_A=q_B=1/2, m_A=m_B=1/3, k_{A,B}= k_{B,A}=1/2,k_{A,AB}= k_{B,AB}=1/4,P_{0}(0) = 10^8\frac{\mathrm{CFU}}{\mathrm{ml}}, P_{A}(0) = 10^3\frac{\mathrm{CFU}}{\mathrm{ml}}, P_B = 10^8\frac{\mathrm{CFU}}{\mathrm{ml}}$, in A: $\rho_{A} = 0.85\mathrm{h}^{-1}, \rho_{AB} = 0.85\mathrm{h}^{-1}$, in B: $\rho_{A} = 0.80\mathrm{h}^{-1}, \rho_{AB} = 0.85\mathrm{h}^{-1}$,in C: $\rho_{A} = 0.85\mathrm{h}^{-1}, \rho_{AB} = 0.9\mathrm{h}^{-1}$,in D: $\rho_{A} = 0.9\mathrm{h}^{-1}, \rho_{AB} = 0.8\mathrm{h}^{-1}$.} \label{fig:coinfection}
\end{figure}

As the number of populations in our model increases, visualizing them by plotting all individual populations becomes more challenging. In such cases, presenting the data through summary statistics may be a more effective approach. However, for the two-plasmid case described above, we can still effectively visualize all populations in a single graph and interpret their dynamics. Here, we will investigate the scenario, where plasmid B is initially present at equal levels to plasmid-free cells, while plasmid A is introduced from rare (i.e. at very low numbers) into the population. If all plasmid states ($P_A, P_B, P_{AB}$) share equal fitness effects (growth costs) and conjugation rates (and we ignore co-conjugation for simplicity), we observe that plasmid A fails to invade the population, despite persisting at low numbers (Fig. \ref{fig:coinfection}A). Similarly, cells carrying both plasmids will remain in the population, but at low abundance.
As we would expect, when plasmid A incurs a higher cost than plasmid B or the co-infected state, both plasmid A and the co-infected state will eventually go extinct over longer time scales (Fig. \ref{fig:coinfection}B).

So far, we have assumed that co-infected cells have the same fitness as cells carrying plasmid B, which is unlikely in natural settings. The co-infected state is particularly interesting because the fitness effects of both plasmids combine in ways that are not immediately apparent. While the fitness of the co-infected state could follow a multiplicative or additive expectation, interactions between plasmids, such as compensatory effects, may result in a net fitness that is either higher or lower than expected from the individual plasmid effects (epistasis).

Understanding the fitness of the co-infected cells is crucial, as beneficial co-infection states can facilitate the invasion of neutral plasmids (those with fitness equal to plasmid B) (Fig. \ref{fig:coinfection}C). Conversely, a costly co-infected state can hinder the invasion of a beneficial plasmid into the population (Fig. \ref{fig:coinfection}D). This occurs because the co-infected state contributes more to the overall fitness of plasmid A than the singly-infected state in a population dominated by plasmid B.

\section{Notes} 
\label{sec:notes}

\subsubsection*{Testing the model}
Models require verification, much like wet lab experiments require controls. To make sure that the model is implemented correctly and its output makes sense, it is good practice to do some sanity checks on the simulated outcomes. A common test would be to set certain parameters to zero, e.g. the growth rate of a particular population or the conjugation rate, and make sure the expected results are recovered, e.g. absence of that particular population. It is essential to visualize some of the raw simulations (population sizes over time) before proceeding to summary statistics (e.g. final population size in each compartment). 

Once one has used the model to identify some interesting predictions, it is also important to test the sensitivity of those predictions to the modeling decisions that were made. Such sensitivity analyses typically involve re-running the same analysis for a range of different parameter values, to see if the interesting behavior is observed across a broader range. In addition, one may want to relax some assumptions of the model structure, to see if the observations continue to hold. This could, for instance, include changing the cost from additive to multiplicative or allowing parameters to depend on growth in some non-trivial way. 




\subsubsection*{Further reading}
Previous guidelines on modeling plasmid dynamics were mainly aimed at readers with a modeling background, looking to get into plasmid biology. These works might nonetheless be of interest for biology-minded readers to obtain a sense of the different plasmid model types that are used to study the ecology and evolution of plasmids.
\begin{itemize}
    \item “Mathematical models of plasmid population dynamics” by Hernández-Beltrán et al.~\cite{hernandez2021}
    \item “A mathematician’s guide to plasmids: an introduction to plasmid biology for modellers” by Dewan and Uecker~\cite{dewan2023}
\end{itemize}

To learn more about compartmental models in general, we recommend introductory books on Infectious disease modeling, such as \textit{Keeling and Rohani}~\cite{keeling2008}.

To learn more about using R for data science, we recommend \textit{Wickham and Grolemund 2017} \cite{wickham2017}.

An introduction to model fitting can be found in Chapter 6 of \textit{Klipp et al. 2016} \cite{klipp2016systems}.

\section{Acknowledgements}
The authors thank Drew Thornley, Kirsten Lim and Taoran Fu for feedback on this chapter. JSH was funded by the Human Frontier Science Program (HFSP) Postdoctoral Fellowship LT0045/2023-L. CI was supported by the Wellcome Trust [225565/Z/22/Z]. The authors used ChatGPT (versions 3.5 and 4o) and Copilot to assist in checking and refining the writing. 

\section{Declaration of interests}
The authors declare no competing interests.\\

\printbibliography

\newpage
\section{Example code} \label{full_R_code} 
\subsection{One plasmid infection model}

\begin{lstlisting}[language=R]
    
    
 require(deSolve)   # install / load required packages

####################################
# Define a function for the one plasmid infection model

plasmid.infection.model <-
  function(sA){
 
## Parameter definition
 pars=list(
      rho_0=1, # growth rate of the plasmid recipient
      rho_A=0.85, # growth rate of the plasmid A carrying cell (including the 
      #cost of the plasmid)
      gamma_0=0.1, #0 # death rate of the plasmid recipient
      gamma_A=0.1, #0 # death rate of the plasmid A carrying cell
      K=10^9, # carrying capacity
      beta_A=1*10^-9, # conjugation rate of plasmid A
      s_A=sA#0.1 #0 # segregation loss rate of plasmid A
    )
    
    
## DE model definition
de_model <- function(t,y,p){
  
 # y[which(y<=1/10^12)]=0  # making sure that unrealistically small 
  # populations are set to 0
  
  with(as.list(c(y,p)),{
    
    T <- sum(c(P_0,P_A))  # total population density
    dP_0=P_0*(rho_0*(1-T/K)-gamma_0-beta_A*P_A)+(1-T/K)*(rho_A*s_A*P_A)
    dP_A=P_A*(rho_A*(1-s_A)*(1-T/K)-gamma_A+beta_A*P_0)
    
    list(c(dP_0,dP_A))
  })
} 

## Simulation condition definition: 
inits <- c(P_0 = 10^5, P_A = 10^5) # initial values of population variables
times <- seq(0, 50, 1)  # time points for simulations 
 
## Solving the DE model:
output <-as.data.frame(lsoda(inits, times, de_model, pars))

return(output) 
}

#######################################

## Run the model
sim <- plasmid.infection.model(sA=0.1)

P_T = sapply(1:length(sim$P_0), function(l) sim$P_0[l]+sim$P_A[l]) # total 
# population size at each time point
plot(sim$time,sim$P_0, ylim = c(10^3,max(P_T)), type='l', log = 'y', 
     xlab='time (h)', ylab='population size (CFU/ml)', col='#D55E00', lwd =3, 
     main = 'with segregation')  # plot P_0
lines(sim$time,sim$P_A, col='#56B4E9', lwd = 3) # add a line for P_A
lines(sim$time,P_T, col='black', lwd =3) # plot total population density
legend("bottomright", legend = c('P_0','P_A','T'), lwd=2, cex=.8, col=c('#D55E00',
      '#56B4E9', 'black'))

\end{lstlisting}


\newpage
\subsection{Two plasmid co-infection model}


\begin{lstlisting}[language=R]


require(deSolve)   # install / load required packages

####################################
# Define a function for the plasmid coinfection model

plasmid.coinfection.model <-
  function(rhoA,rhoAB){
    
    ## Parameter definition
    pars=list(
      rho_0=1, # growth rate of the plasmid recipient
      rho_A=rhoA, #0.8,  #0.9 #0.85 # 0.8 # growth rate of the plasmid A carrying cell 
      rho_B=0.85,  # growth rate of the plasmid B carrying cell 
      rho_AB=rhoAB, #0.85, #0.9 #0.85 # 0.8 # growth rate of the plasmid A and B carrying cell 
      gamma_0=0.1, #0 # death rate of the plasmid recipient
      gamma_A=0.1, #0 # death rate of the plasmid A carrying cell
      gamma_B=0.1, #0 # death rate of the plasmid B carrying cell
      gamma_AB=0.1, #0 # death rate of the plasmid A and B carrying cell
      K=10^9, # carrying capacity
      beta_A=10^-9, # conjugation rate of plasmid A
      beta_B=10^-9,  # conjugation rate of plasmid B
      beta_AB=0, # co-conjugation rate of plasmids A and B
      q_A=1/2, # relative transmissibility of plasmid A from co-infected cells
      q_B=1/2, # relative transmissibility of plasmid B from co-infected cells
      s_A=0.1, #0 # segregation loss rate of plasmid A
      s_B=0.1, #0 # segregation loss rate of plasmid B
      s_AB=0.1, #0 # segregation loss rate of plasmids A and B
      m_A=1/3, #0 # partial loss rate of plasmid A from coinfected cells
      m_B=1/3, #0 # partial loss rate of plasmid B from coinfected cells
      k_AAB=1/4, # probability of plasmid A displacing resident plasmids A and B
      k_BAB=1/4, # probability of plasmid B displacing resident plasmids A and B
      k_AB=1/2,  # susceptibility of cells with plasmid B to infection with 
      #plasmid A
      k_BA=1/2,  # susceptibility of cells with plasmid A to infection with 
      #plasmid B
      g_A=0,  # probability of co-transmission leading to co-infection of 
      #cells infected with plasmid A
      g_B=0  # probability of co-transmission leading to co-infection of 
      #cells infected with plasmid B
    )
    
    
    ## DE model definition
    de_model <- function(t,y,p){
      
      # y[which(y<=1/10^12)]=0  # making sure that unrealistically small 
      # populations are set to 0
      
      with(as.list(c(y,p)),{
        
      T <- sum(c(P_0,P_A, P_B, P_AB))  # total population density
      lambda_A = beta_A*(P_A+q_A*P_AB)
      lambda_B = beta_B*(P_B+q_B*P_AB)
      
      dP_0=P_0*(rho_0*(1-T/K)-gamma_0-lambda_A-lambda_B-beta_AB*P_AB)+
      (1-T/K)*(rho_A*s_A*P_A+rho_B*s_B*P_B+rho_AB*s_AB*P_AB)
      
      dP_A=P_A*(rho_A*(1-s_A)*(1-T/K)-gamma_A-k_BA*(lambda_B+g_A*beta_AB*P_AB))+
      lambda_A*(P_0+k_AAB*P_AB)+m_A*rho_AB*(1-s_AB)*(1-T/K)*P_AB
      
      dP_B=P_B*(rho_B*(1-s_B)*(1-T/K)-gamma_B-k_AB*(lambda_A+g_B*beta_AB*P_AB))+
      lambda_B*(P_0+k_BAB*P_AB)+m_B*rho_AB*(1-s_AB)*(1-T/K)*P_AB
      
      dP_AB=P_AB*(rho_AB*(1-s_AB)*(1-m_A-m_B)*(1-T/K)-gamma_AB +
      beta_AB*(P_0+g_A*k_BA*P_A+g_B*k_AB*P_B)-lambda_A*k_AAB-lambda_B*k_BAB)+
      lambda_B*k_BA*P_A+lambda_A*k_AB*P_B
      
      
        list(c(dP_0,dP_A, dP_B, dP_AB))
      })
    } 
    
    ## Simulation condition definition: 
    inits <- c(P_0 = 1*10^8, P_A = 10^3, P_B = 1*10^8, P_AB = 0) # initial population values
    times <- seq(0, 5000, 1)  # time points for simulations 
    
    ## Solving the DE model:
    output <-as.data.frame(lsoda(inits, times, de_model, pars))
    
    return(output) 
  }

#######################################

## Run the model
sim <- plasmid.coinfection.model(rhoA=0.8, rhoAB=0.85)

P_T = sapply(1:length(sim$P_0), function(l) sim$P_0[l]+sim$P_A[l]+sim$P_B[l]
             +sim$P_AB[l]) # total 
# population size at each time point
plot(sim$time,sim$P_0, ylim = c(10^3,max(P_T)), type='l', log = 'y',
     xlab='time (h)', ylab='population size (CFU/ml)', col='#D55E00', lwd =3, 
     main = 'invasion of beneficial plasmid - costly coinfection')  
lines(sim$time,sim$P_A, col='#56B4E9', lwd = 3) # add a line for P_A
lines(sim$time,sim$P_B, col='#009E73', lwd = 3) # add a line for P_A
lines(sim$time,sim$P_AB, col='#F0E442', lwd = 3) # add a line for P_A
lines(sim$time,P_T, col='black', lwd =3) # plot total population density
legend("bottomright", legend = c('P_0','P_A','P_B', 'P_AB','T'), lwd=2, cex=.8, 
       col=c('#D55E00', '#56B4E9', '#009E73','#F0E442', 'black'))







\end{lstlisting}


\end{document}